\documentclass[journal,onecolumn,twoside,a4paper]{IEEEtran}
\usepackage{xcolor}

\usepackage{url}
\usepackage{ifthen}
\usepackage{cite}
\usepackage[cmex10]{amsmath} 
\usepackage{enumitem}
\usepackage{revsymb}

\newtheorem{theorem}{Theorem}
\newtheorem{lemma}[theorem]{Lemma}
\newtheorem{definition}[theorem]{Definition}

\newtheorem{example}[theorem]{Example}
\newtheorem{remark}[theorem]{Remark}

\newcommand\qed{\hfill\IEEEQED}

\newlength{\blank}
\newenvironment{proof-of}[1][{\hspace{-\blank}}]{{\medskip\noindent\textit{Proof~{#1}.\ }}}{\hfill\IEEEQED}

\usepackage{csquotes}
\usepackage{bbm}
\usepackage{mathtools}
\usepackage{comment}
\usepackage{physics}
\usepackage{amssymb}
\renewcommand{\Tr}{{\operatorname{Tr}}}
\newcommand{\id}{\operatorname{id}}
\newcommand{\EE}{\mathbb{E}}
\newcommand{\RR}{\mathbb{R}}
\newcommand{\cX}{\mathcal{X}}
\newcommand{\cT}{\mathcal{T}}

\newcommand{\1}{\openone}
\newcommand{\proj}[1]{|#1\rangle\!\langle #1|}
\newcommand{\ox}{\otimes}

\begin{document}

\title{Multi-User Distillation of Common Randomness and Entanglement from Quantum States} 

\author{Farzin~Salek and Andreas~Winter%
\thanks{The authors are with F\'{\i}sica Te\`{o}rica: Informaci\'{o} i Fen\`{o}mens Qu\`{a}ntics, Departament de F\'{\i}sica, Universitat Aut\`{o}noma de Barcelona, 08193 Bellaterra (Barcelona), Spain.}%
\thanks{FS is with Departamento de Teor\'{i}a de la Se\~{n}al y Comunicaciones (TSC), Universitat Polit\`{e}cnica de Catalunya, Barcelona, Spain. Email: farzin.salek@gmail.com}%
\thanks{AW is with ICREA---Instituci\'o Catalana de Recerca i Estudis Avan\c{c}ats, Pg.~Lluis Companys, 23, 08010 Barcelona, Spain. Email: andreas.winter@uab.cat}%
\thanks{Dated: 10 August 2020. A short version of this work has been presented at ISIT 2020 \cite{GHZ:ISIT}.}}

\maketitle

\begin{abstract}
We construct new protocols for the 
tasks of converting noisy multipartite quantum correlations into noiseless 
classical and quantum ones using local operations and classical communications (LOCC). 
For the former, known as common randomness (CR) distillation, two new lower bounds on 
the ``distillable common randomness'', an operational measure of the total 
genuine (classical) correlations in a quantum state, are obtained. 
Our proof relies on a generalization of communication for omniscience (CO) 
[Csisz\'{a}r and Narayan, IEEE Trans. Inf. Theory 50:3047-3061, 2004]. Our 
contribution here is a novel simultaneous decoder for the compression of correlated
classical sources by random binning with quantum side information at the decoder. 
For the latter, we derive two new lower bounds on the rate at which 
Greenberger-Horne-Zeilinger (GHZ) states can be asymptotically distilled 
from any given pure state under LOCC. 
Our approach consists in ``making coherent'' the proposed CR distillation 
protocols and recycling of resources [Devetak \emph{et al.} 
IEEE Trans. Inf. Theory 54(10):4587-4618, 2008]. 
The first lower bound is identical to a recent result by Vrana and Christandl
[IEEE Trans. Inf. Theory 65(9):5945-5958, 2019], which is based on a combinatorial 
method to achieve the same rate. Our second lower bound generalises and 
improves upon this result, and unifies a number of other known lower bounds on 
GHZ distillation. 
%
\end{abstract}

\section{Introduction and preliminaries}
Interconversion between various resources is one of the big ongoing programs 
of quantum and classical information theory for a considerable time \cite{4626055}. 
Within that broad class of questions, the transformations of 
multipartite quantum states into other forms has provided considerable 
inspiration. A particularly prototypical example of 
this is bipartite entanglement of pure states: in the asymptotic setting 
of many copies of a pure state, not only can each pure state $\ket{\psi}^{AB}$ be 
converted to EPR states $\ket{\phi} = \frac{1}{\sqrt{2}}(\ket{0}\ket{0}+\ket{1}\ket{1})$ 
at rate $E(\psi)=S(A)_\psi$ by local operations and classical communication
(LOCC), where $S(A)_{\rho}=-\Tr\rho^{A}\log\rho^{A}$ is the von Neumann entropy 
of the reduced state of a quantum state $\rho^{AB}$, the same rate governs 
the reverse transformation from $\phi$ to $\psi$ \cite{PhysRevA.53.2046}. 
The story is far less satisfying for mixed states \cite{HHHH-review},
nevertheless this raised certain expectations for multipartite pure states: 
while it is clear that there cannot be a single ``gold standard'' like the 
EPR state in the bipartite setting -- as EPR states between any 
pair of $m$ parties are inequivalent to EPR 
states between any other pair --, the question arose whether there is a ``minimal 
reversible entanglement generating set'' (MREGS) \cite{BPRST}. In the latter
paper, it was shown that for $m\geq 4$ parties, also the GHZ state
$\ket{\Gamma_{m}} = \frac{1}{\sqrt{2}}(\ket{0}^{\otimes m}+\ket{1}^{\otimes m})$
needs to be part of an MREGS, and in \cite{LPSW} this was extended to $m=3$. 
Since then, increasing lower bounds on the size of an MREGS have been proved, 
and it is conceivable that any MREGS is infinitely large. 
For a broad overview over the history and state of the art in multipartite 
entanglement, see the review \cite{WalterGrossEisert:review}. 

In any case, the frustrated hope of the MREGS programme has made researchers reevaluate 
what we actually want from our theory of state conversions. One big component, rather 
than a universal normal form, is knowledge how, and how efficiently, to transform a 
given $m$-partite pure state $\ket{\psi}^{A_1\ldots A_m}$ into a specific desired 
target state. 
In the multipartite setting, this presents a problem of choice. 
There seem to be at least two canonical options: 
first, aim for EPR states between designated pairs of parties, and second, 
an $m$-party GHZ state. The first problem has an elegant solution, based on 
quantum state merging \cite{merging-Nature}. If EPR states are to be distilled
between a specific pair of parties, say $i$ and $j$, then the optimal rate (capacity) 
is the following number \cite{merging-CMP}:
\begin{equation}
  \label{eq:EPR-pairwise}
  C_{\text{EPR}(i:j)}(\psi) = \min_I S(A_I)_\psi \text{ s.t. }  i\in I\subseteq[m]\setminus j.
\end{equation}
If we want to distill EPR states between different pairs of parties simultaneously, 
there are partial results, for example outer rate bounds from the subgroup 
entropies, all of which are monotones \cite[Lemma 1 \& Thm. 2]{BPRST}, 
i.e. each $S(A_I)$, for $I\subseteq [m]$, is a monotone under asymptotic LOCC;
furthermore, \cite{LPSW} gives asymptotic monotones for certain state 
conversions based on the quantum relative entropy. 
And there is 
the ``entanglement combing'' protocol that yields EPR pairs between a single party 
and each of the other $m-1$ \cite{PhysRevLett.103.220501}. 
These tasks of creating pairwise (EPP type) entanglement between nodes, assisted 
by the others, is very much tied to the objectives of the so-called
quantum internet \cite{DrKimble,Godwin}.
As for GHZ distillation, also this is evidently relevant for the quantum 
internet, but has received considerably less attention; we review some of the
relevant prior work below. 

In the present paper, we address this second-tier type of question via a 
two-pronged strategy.
The first resource conversion we study is the task of converting 
noisy multipartite quantum correlations, i.e. an $m$-partite quantum state ($m\geq2$), 
into noiseless $m$-partite classical correlations, i.e. common randomness (CR), 
under local operations and classical communications (LOCC). Intuitively, CR is 
a random variable that is uniformly distributed and known to all $m$ parties. 
It is known that distillation of CR without additional classical communication 
is generically impossible \cite{651026}. On the other hand, since classical 
communication and CR are not ``orthogonal'' resources, allowing free classical 
communications is not appropriate, because it can be used to create unlimited CR. 
However, one can consider two interesting directions: 
imposing a secrecy requirement on CR, or limiting the classical communication. 
In this paper, we are concerned with the second direction; the first one, known 
as key distillation, was studied by Maurer \cite{Maurer:key}, 
Ahlswede and Csisz\'ar \cite{243431} and its 
quantum generalization in \cite{2005RSPSA.461..207D}. 
The problem of distilling CR from two correlated random variables under one-way 
classical communication of $R$ bits per source observation was studied by 
Ahlswede and Csisz\'ar \cite{651026} (see the paper for other models). 
Subsequently, their model was generalized in \cite{1362905}, 
introducing the \emph{distillable CR}, the amount of CR generated in excess 
of the consumed classical communication. When the classical communication is 
one-way, the distillable CR is still an (asymmetric) measure 
of the total classical correlations in the state \cite{HendersonVedral}.  
For a recent review of multi-party key distillation see \cite{MGKB}.
 
In Section \ref{sec:CR}
we prove two lower bounds on the distillable CR from multipartite mixed quantum states.
We do this by offering a generalization of a result in multi-terminal 
distributed lossless source coding and secret key agreement due to Csisz\'{a}r and 
Narayan \cite{1362897} known as \emph{communication for omniscience} (CO). 
There, $m$ parties observe a correlated discrete memoryless multiple source 
$X_{[m]} = (X_{1},\ldots,X_{m})$, the $i$-th node obtaining
$X_{i}$. The nodes are allowed to communicate interactively over a public noiseless 
broadcast channel so that at the end they attain omniscience: each node reconstructs 
the whole vector of observations $X_{[m]}$. 
The objective is to minimise the overall communication to achieve this goal. 
%
%
We first apply the main result of \cite{1362897} to the outcomes of 
local measurements on an $m$-partite quantum states, and then generalize this 
result to partial measurements, modelled as instruments, such that  
each party not only has a classical information $X_i$ but also a quantum 
register $A_i'$ containing containing correlated quantum side information. 
It uses a novel random binning coding and decoding strategy for 
the problem of correlated source compression with quantum side information 
at the decoder, presented in a concise way in the Appendix. 
The reason for the secrecy rate being exactly the difference between 
the entropy of $X_{[m]}$ and the total communication rate $R_{\text{CO}}$
is that this is attained by privacy amplification. We note that the same rate is 
also an achievable rate for the distillable CR by the recycling of resources idea; 
for more on their relation see \cite{8328871}.

Our second problem concerns converting multipartite quantum correlation 
into noiseless quantum correlation, i.e. the so-called entanglement distillation 
task (Section \ref{sec:GHZ}). 
The theory of asymptotic manipulation of multipartite entanglement is 
very complex, even in the pure state case a simple theory as is known for bipartite 
pure states, is probably forever beyond reach; for mixed states, already the 
bipartite case defies complete analysis, so much so that it is even open 
whether there are bound entangled states with non-positive partial transpose (NPT). 
For these reasons, for the task of entanglement distillation, we focus on the 
Greenberger-Horne-Zeilinger (GHZ) distillation problem, and on pure initial states.
Very little previous work has concerned itself with the asymptotic rate of GHZ
distillation, despite such states being evidently useful for cryptography \cite{PhysRevA.59.1829}.
The important exceptions are Smolin \emph{et al.} \cite{PhysRevA.72.052317}, 
Fortescue and Lo \cite{FortescueLo} and Streltsov \emph{et al.} \cite{Streltsov-et-al};
furthermore \cite{BFG} for stabilizer states and exact distillation. In \cite{Takeoka-et-al},
general upper rate bounds are established that go beyond the entropy and relative 
entropy bounds from \cite{BPRST,LPSW}.

Motivated by the recent paper \cite{8678669}, which treats the distillation of 
multipartite GHZ states from many copies of a given multipartite pure state and 
presents an achievable rate based on a combinatorial construction, we realised 
that the same rate can be obtained and improved using off-the-shelf techniques of 
quantum Shannon theory from the early 2000s, namely the coherification of protocols 
for CR distillation. 
%
%
%
%
The first lower bound reproduces the result of Vrana and Christandl \cite{8678669}, 
and the second protocol improves upon this lower bound. 
To the best of our knowledge it is the best available bound,
subsuming a number of other previous results. 

In Section \ref{sec:end} we conclude the paper with a brief discussion and three 
example states for which we can compare our new and the old achievable rates of 
GHZ distillation. 


\medskip
\textbf{Notation.} Capital letters $X$, $Y$, etc. denote random variables, whose
realizations and the alphabets are shown by the corresponding small and calligraphic 
letters, respectively: $X=x \in \mathcal{X}$.
Quantum systems $A$, $B$, etc. are associated with (finite-dimensional) 
Hilbert spaces $A$, $B$, etc. whose dimensions are denoted by $|A|$, $|B|$, etc. 
Multipartite systems $AB\ldots Z$ are described by tensor product Hilbert space 
$A\otimes B\otimes \cdots \otimes Z$. We identify states with their density operators 
and use superscripts to denote the systems on which the mathematical objects are defined. 
For any positive integer $m$, we use the notation $[m]=\{1,...,m\}$. 
For conciseness, we denote the tuple $(X_{1},..,X_{m})$ by $X_{[m]}$.
More generally, for a set $L$, we write $X_{L}=(X_{i}: i\in L)$. 
Throughout the paper, $\log$ denotes by default the binary logarithm. 

Beyond the von Neumann entropy of a state, we also use 
the conditional von Neumann entropy of a bipartite state $\rho^{AB}$, 
defined as $S(A|B)=S(AB)-S(B)$, and the quantum mutual information
$I(A:B) = S(A)+S(B)-S(AB)$. 
For classical systems (random variables), the von Neumann entropy 
reduces to the Shannon entropy, denoted $H(X)$.

\section{Common randomness distillation and omniscience}
\label{sec:CR}
We shall consider \emph{common randomness distillation} (in the source 
model). This means that we have $m$ spatially separated parties sharing 
$n \gg 1$ copies of an $m$-partite quantum state $\rho^{A_{1}\ldots A_{m}}$, i.e. party 
$i$ holds the subsystem $A_i^n$. All parties can communicate to each other 
through a public noiseless classical broadcast channel of unlimited capacity.
%
The following definition is a generalization of the bipartite case in \cite{20.500.11850/153605}.

\medskip
\begin{definition}[Common randomness distillation protocol]
Let $\rho$ be a state on $A_{[m]}=A_{1}\otimes\cdots\otimes{A_{m}}$,
and consider the initial state $\rho^{\otimes n}$.
Let $r$ be the total number of rounds; 
for $i\in[m]$, let $B_{i}$ be a local quantum system used by party $i$ to store 
quantum information, originally in state $\ketbra{0}$;
for $j\in [r]$, let $U^{j}_{i_j}$ be classical systems to store the classical
communication of party $i_j$ after round $j$.

\setlist[description]{font=\textbf\space}
\begin{description}
  \item [Step 1)] Terminal $i_1\in[m]$ applies the completely positive instrument
        \begin{align*}
          \Phi_{i_1}^{1}:A_{i_1}^{n}\otimes B_{i_1} 
                           \rightarrow A_{i_1}^{n}\otimes B_{i_1}\otimes U^{1}_{i_1},
        \end{align*}
        and broadcasts $U^{1}_{i_1}$ to the other parties. This means that the shared state 
        $\rho^{\otimes n}$ is mapped to the state
        \begin{align*}
          \rho^{(1)} 
            &= \sum_u (\id_{A_{[m]\setminus i_{1}}^n}\otimes\Phi_{i_1}^{1}(u))(\rho^{\otimes n}\otimes\ketbra{0}^{B_{i_1}}) 
                      \otimes \ketbra{0}^{B_{[m]\setminus i_{1}}} \otimes \ketbra{u}^{U^{1}_{i_1}}
        \end{align*}
        on $A_{[m]}^{n} \otimes B_{[m]}\otimes U_{i_{1}}^{1}$.

  \item [Step j)] Terminal $i_j\in[m]$ applies a completely positive map
        \begin{align*}
          \Phi_{i_{j}}^{j}:A_{i_{j}}^{n}\otimes B_{i_{j}}\otimes U^{[j-1]} 
                   \rightarrow A_{i_{j}}^{n}\otimes B_{i_{j}}\otimes U^{[j]},
        \end{align*} 
        where we use the shorthand $U^{[j-1]}=U^1_{i_1} U^2_{i_2} \ldots U^{j-1}_{i_{j-1}}$,
        and broadcasts $U^{j}_{i_j}$ to the rest of the parties. 
        This maps the previous state $\rho^{(j-1)}$ to the new state $\rho^{(j)}$ 
        on $A_{[m]}^{n}\otimes B_{[m]}\otimes U^{[j]}$.

  \item [Step r+1)] After the last communication, each party $i\in[m]$, 
        measures its systems by means of a POVM on 
        $A_{i}^{n}\otimes B_{i}\otimes U^{[r]}$ and indexed by $\{1,\ldots,|V|\}$,
        giving rise to a random variable $V_{i}$ with distribution $p_{i}(v)$. 
\end{description}
Let $R_{i}$ denote the total rate of classical communication by the $i$-th party. 
\end{definition}

\medskip
\begin{remark}
This CR distillation protocol is a general LOCC procedure, in which we
explicitly keep track of the classical communication. 
\end{remark}


\medskip
\begin{definition}
A number $R=\frac{1}{n}\log|V|-\sum_{i=1}^{m}R_{i}$ will be called an 
achievable distillable CR rate for common randomness distillation if for every 
$\varepsilon>0$ and sufficiently large $n$, there exists a common randomness 
distillation protocol where the total communication of party $i$ is 
bounded by $nR_i$ bits, such that $\{V_{i}\}_{i=1}^{m}$ satisfy 
\begin{align}
\text{Pr}\{V_{1} = \ldots = V_{m}\}                                            &\geq 1-\varepsilon,\\
\frac12 \|p_1-u_V\|_1 = \frac{1}{2}\sum_{v}\left|p_{1}(v)-\frac{1}{|V|}\right| &\leq \varepsilon,
\end{align}
where $u_V$ denotes the uniform distribution. 
The maximal achievable rate for distillable CR is called the 
\emph{distillable CR capacity} $D_{CR}(\rho)$. 
\end{definition}

\medskip
Now, we prove two achievability results for the distillable CR rate, 
and in the next section two achievability results for the distillable GHZ rate, 
all based on a subclass of protocols with ``non-interactive communication'',
which are called this way because each party broadcasts only one 
message to all others that depends only on their own local state. 
The proof of the distillable CR results is based on our generalization of the 
communication for omniscience (CO) \cite{1362897}. We present two protocols for our 
achievability bounds. The first protocol uses full local measurements and communication;
the second uses instruments that initially turn the state into a 
classical-quantum state, and then generalizes the first.

\medskip
\begin{theorem}
\label{lowercr1}
Let $\rho^{A_{1}\ldots A_{m}}$ be a quantum state and let 
$\{M^{i}_{x_i}\}_{x_i\in\mathcal{X}_{i}}$ denote a POVM used by party $i$.
Define $p(x_{[m]})$ as the joint distribution of $m$ random variables $X_i$ 
recording the measurement outcomes on $\rho$:
\[
  p(x_1,\ldots,x_m) = \Tr\rho(M^{1}_{x_1}\otimes\cdots\otimes M^{m}_{x_m}).
\] 
The following is an achievable rate for the distillable CR:
\begin{align*}
  R = H(X_{[m]})-R_{\text{CO}}^{c},
\end{align*}
where 
$R_{\text{CO}}^{c}={\displaystyle{\min_{R_{[m]}\in\mathcal{R}_{c}}}} \sum_{i=1}^{m} R_{i}$,
and $\mathcal{R}_{c}$ is the rate region of tuples $R_{[m]}=(R_1,\ldots,R_m)$
given as follows:
\begin{align*}
  \forall L\subsetneq [m] \quad \sum_{j\in L} R_{j}\geq H(X_{L}|X_{[m]\setminus L}).
\end{align*}
\end{theorem} 

\begin{IEEEproof}
This really is an instance of the results of Csisz\'ar and Narayan \cite{1362897},
who prove precisely that for the RVs $X_1,\ldots,X_m$, the set $\mathcal{R}_{c}$
is precisely the rate region of communication for omniscience, 
i.e. protocols at the end of which 
all users know $X_{[m]}$ up to arbitrarily small error probability. This shows 
that $R = H(X_{[m]})-R_{\text{CO}}^{c}$ is an achievable rate for distillable CR. 
Incidentally, in \cite{1362897} it is actually shown to be the optimal CR rate 
for the given RVs. However, this is of less relevance for us, as different 
choices of local measurements lead to different tuples of RVs.
The theorem is also a special case of Theorem \ref{lowercr2} below. 
The basic idea of the coding procedure, referred to as \emph{random binning}, is not
much different than that of hash functions.
Each classical sequence obtained from the local measurements is randomly and uniformly 
assigned a bin index; 
if the number of bins (the range of the hash function) is large enough compared to the 
jointly entropy-typical sets, a randomly selected mapping of classical sequences will 
suffer a collision with small probability. This means that the
classical information can be extracted from their index set with high probability. 

In detail, the $i$-th party assigns each sequence $x_{i}^{n}\in\mathcal{X}_{i}^{n}$ 
to one of $2^{nR_{i}}$ bins; 
all parties broadcast the bin index associated to their obtained sequence, 
$(\mu_{1},\ldots,\mu_{m})\in[M_{1}]\times\cdots\times [M_{m}]$,
to the other parties.
Then, the parties use joint typicality decoding to extract the sequences of other 
parties from their local information and $\mu_{[m]}$. That is, having received 
$\mu_{[m]\setminus i}$, the $i$-th party looks into the bins indexed $\mu_{[m]\setminus i}$ 
to find a unique tuple $\hat{x}_{[m]\setminus i}^{n}$ that is jointly typical 
with their observed $x_{i}^{n}$. 
An error occurs when one of the following events happen: 
the obtained sequence of tuples $x_{[m]}^n$ is not typical, 
or there is no jointly typical 
sequence $\hat{x}_{[m]\setminus i}^{n}x_i^n$, or there are two different 
jointly typical candidates $\hat{x}_{[m]\setminus i}^{n}x_i^n$ and 
$\check{x}_{[m]\setminus i}^{n}x_i^n$ in the correct bins. 
These in fact are the same conditions as for correct decodability in the Slepian-Wolf 
problem \cite[Ch.~15.4]{CoverThomas}, in the special case that $R_i = H(X_i) + \delta$, 
for some $\delta > 0$. The analysis there shows that the error probability goes 
indeed to zero, with high probability for a randomly chosen binning strategy, 
if for all $L\subseteq [m]\setminus i$ it holds 
$\sum_{j\in L} R_{j} \geq H(X_{L}|X_{[m]\setminus L}) + \delta$,
for some $\delta>0$.

As $\mathcal{R}_{c}$ consists of the rate tuples satisfying 
these conditions for all $i\in[m]$, it means that then all parties can 
decode $x_{[m]}^n$ with high probability correctly, as $n\rightarrow \infty$. 
\end{IEEEproof}

\medskip
\begin{theorem}
\label{lowercr2}
Let $\rho^{A_{1}\ldots A_{m}}$ be a quantum state and let
$\mathcal{E}^{i}: A_{i}\rightarrow A_{i}'\otimes X_{i}$ be an instrument used by party $i$,
with quantum registers $A_i'$ and classical registers $X_i$. 
Define $\omega^{X_{1}A_{1}'\ldots X_{m}A_{m}'}$ as the cq-state after applying 
the local instruments: 
\[\begin{split}
  \omega^{X_{1}A_{1}'\ldots X_{m}A_{m}'} 
                  &= (\mathcal{E}^{1}\otimes\cdots\otimes\mathcal{E}^{m})\rho \\
                  &= \sum_{x_{[m]}} \ketbra{x_{[m]}}^{X_{[m]}} 
                                      \otimes 
                                    (\mathcal{E}^{1}_{x_1}\otimes\cdots\otimes\mathcal{E}^{m}_{x_m})\rho.
\end{split}\]
The following is an achievable rate for the distillable CR:
\begin{align*}
  R = H(X_{[m]})-R_{\text{CO}}^{cq},
\end{align*}
where 
$R_{\text{CO}}^{cq}=\min_{\substack{R_{[m]}}\in\mathcal{R}_{cq}}\sum_{i=1}^{m} R_{i}$,
and $\mathcal{R}_{cq}$ is the rate region given as follows: 
\begin{align}
  \label{eq:R_cq}
  \forall j\in [m]\ \forall L\subseteq [m]\setminus j \quad 
      \sum_{i\in L} R_{i} \geq S(X_{L}|X_{[m]\setminus L}A_{j}').
\end{align}
\end{theorem}  

\begin{IEEEproof}
Each party $j$ evaluates a function 
$U_j:= f_j(X_j^n) \in \{0,1\}^{nR_j}$
of their input, and broadcasts $U_j$ to all other parties. The objective 
for party $j$ is then, knowing $U_{[m]\setminus j}$, that 
they can decode $X_{[m]}^n$ from $B_j^n := X_j^n{A_j'}^n$ by a suitable measurement. 

Thus it is unsurprising that the answer should be given by a quantum 
version of Slepian-Wolf coding. Indeed, for each fixed $j$, the 
necessity and sufficiency of the rate conditions in Eq. (\ref{eq:R_cq}) 
is proved in \cite[Thm. IV.14 \& Cor. IV.16]{WINTER:PHD}, generalising 
\cite{PhysRevA.68.042301}. 
However, this is not enough because we need a code (i.e. a set of encoding functions,
one for each party) that works for all parties simultaneously, allowing each 
of the to recover $X_{[m]}^n$ for their $A_j'$ and $U_{[m]}$. 
To achieve this, we use random binning: each party $j$ uses a random function 
$F_j:\mathcal{X}_j^n\rightarrow \{0,1\}^{nR_j}$
(to be precise, we draw them independently from $m$ $2$-universal families). 
In the case of classical $A_i'$, it is well-known that this strategy 
works as long as the rate conditions in Eq. (\ref{eq:R_cq}) are satisfied, 
by using a joint typicality decoder, see the proof sketch 
of Theorem \ref{lowercr1}; cf. the discussion of 
Slepian-Wolf data compression in \cite[Ch.~15.4]{CoverThomas}. 
In the general quantum case, joint typicality decoding presents considerable 
technical difficulties, but they were eventually overcome by Sen \cite{Sen:jointly-typical}.

In Lemma \ref{lemma:cq-compression-simultaneous} in the Appendix, we show how to use 
Sen's joint typicality construction to build a joint decoder that achieves
small expected decoding error for party $j$, 
$\EE_{F_{[m]\setminus j}} P_e(j) \leq \epsilon$, for any $\epsilon > 0$
and sufficiently large $n$, if the rates satisfy 
\begin{equation*}
  \forall\ \emptyset\neq I\subset[m]\setminus j\quad 
           \sum_{i\in I} R_i \geq H(X_I|X_{[m]\setminus j\setminus I}B_j) + \delta,
\end{equation*}
where $\delta>0$ is an arbitrary constant. Thus, summing over all $j$,
and recalling $B_j=X_j A_j'$, 
we get $\EE_{F_{[m]}} \bigl(P_e(1)+\ldots+P_e(m)\bigr) \leq m\epsilon$ 
for all sufficiently large $n$, if the rates satisfy
\[
  \forall j\in [m]\ \forall L\subseteq [m]\setminus j \quad 
      \sum_{i\in L} R_{i} \geq S(X_{L}|X_{[m]\setminus L}A_{j}') + \delta.
\]
Since $\epsilon,\delta > 0$ are arbitrary, the claim follows.

This shows that the rate tuples $(R_1,\ldots,R_m) \in \mathcal{R}_{cq}$ 
are all achievable to provide omniscience of the $X_{[m]}^n$ among all $m$ 
parties. Concentrating the randomness in the shared random variables 
into uniform randomness, yielding a rate of $H(X_{[m]})$, and subtracting 
the communication $\sum_i R_i$, completes the proof that $R = H(X_{[m]})-R_{\text{CO}}^{cq}$
is an achievable rate for distillable CR. 
\end{IEEEproof}

\medskip
\begin{remark}
It is easy to see, via the Slepian-Wolf connection made in the above proof, 
that given the cq-state $\omega^{X_{[m]}A_{[m]}'}$, any non-interactive 
protocol to achieve omniscience of $X_{[m]}$, by which party $j$ broadcasts at 
asymptotic rate $R_j$, must necessarily satisfy $(R_1,\ldots,R_m) \in \mathcal{R}_{cq}$.

Indeed, focusing on party $j$ for the moment, for them to be able to 
reconstruct $X_1^n,\ldots,X_{j-1}^n,X_{j+1}^n,\ldots,X_m^n$ using 
$X_j^n{A_j'}^n$ and communications $U_i$ from party $i\in[m]\setminus j$
at rate $R_i$, is precisely the task of correlated classical source coding 
with quantum side information at the decoder \cite{PhysRevA.68.042301,WINTER:PHD}.
For this, the conditions in Eq. (\ref{eq:R_cq}) for the given $j$
are necessary and sufficient. Since they have to hold for all $j$, it
follows that $\mathcal{R}_{cq}$ is precisely the achievable region of 
rates for CO.
\end{remark}

\section{GHZ distillation from pure states}
\label{sec:GHZ}
Now, we move on to using the above results on distillable CR to derive two 
lower bounds for the distillable entanglement from pure quantum states. 
The first, Theorem \ref{Vrana-Christandl}, re-derives the result of \cite{8678669}, 
with a different, information theoretic, proof, 
by making the protocol of Theorem \ref{lowercr1} coherent. 
The second, which improves upon the preceding result, 
is obtained by making the protocol of Theorem \ref{lowercr2} coherent. 
We use lessons learned 
in \cite{1377491,PhysRevLett.93.080501,2005RSPSA.461..207D,PhysRevLett.93.230504}, and 
observations of \cite{PhysRevLett.92.097902} regarding making protocols coherent. 

In short, the first idea of making protocols coherent is that classical symbols $x$ 
become basis states $\ket{x}$ of the Hilbert space. Functions $f:x\rightarrow f(x)$ 
thus induce linear operators on Hilbert space, but only permutations (resp. 
one-to-one functions) are really interesting, since they give rise to unitaries 
(resp. isometries). The second idea is thus to make classical computations first 
reversible, by extending them into one-to-one functions. The last step is to use 
the local decoding operations that exist by the ``classical'' theorems, which
are cptp maps, in the form of their isometric Stinespring dilations \cite{Stinespring}. 
In summary, ``making coherent'' means we can take a classical protocol working on letters 
and turn it into a bunch of unitaries acting as permutations on the basis states, 
and that we can run perfectly well on superpositions.

As in CR distillation, we have $m$ spatially separated parties, now sharing 
$n \gg 1$ copies of an $m$-partite \emph{pure} quantum state $\ket{\psi}^{A_{1} \ldots A_{m}}$, 
i.e. party $i$ holds the subsystem $A_i^n$. All parties can communicate to 
each other through a public noiseless classical broadcast channel of unlimited capacity.

\begin{definition}[GHZ distillation protocol]
The $m$ parties, to convert the state $\psi^{\otimes n}$ to $k$ copies of the GHZ 
state $\ket{\Gamma_{m}}$, 
they perform LOCC channels interactively in $r$ rounds. Let $\sigma^{B_{1}^{k} \ldots B_{m}^{k}}$ 
denote the final state after LOCC channels, where $B_i$ denotes qubit systems. If
\begin{align*}
  \frac{1}{2}\left\|\sigma^{B_{1}^{k} \ldots B_{m}^{k}}-\ketbra{\Gamma_{m}}^{\otimes k}\right\|_{1}
     \leq\varepsilon,
\end{align*}
we call the protocol $\varepsilon$-accurate and the GHZ conversion rate is $k/n$. 
We call a number $R$ an achievable rate for GHZ distillation if there exists an 
$\varepsilon$-accurate protocol with conversion rate $R-\varepsilon$ for all $\varepsilon>0$. 
The supremum of all achievable rates is the GHZ distillation capacity, $C_{\text{GHZ}}(\psi)$.
\end{definition}

\medskip 
At the time of writing, there is no formula known for $C_{\text{GHZ}}(\psi)$ for
a general state, however various protocols (giving lower bounds) and upper 
bounds have been developed. Regarding the latter, this involves finding LOCC 
monotones that have certain requisite additivity and continuity properties.
For example, in \cite[Lemma~1 \& Thm.~2]{BPRST} it was shown that for multipartite
pure state transformation, all the $S(A_I)_\psi$, $I\subset [m]$, are such 
monotones, thus limiting the conversion rate for any target state. In the 
case of a GHZ state, which has $S(A_I)_{\Gamma_m} = 1$ for all $\emptyset \neq I \subsetneq [m]$,
this leads to
\begin{equation}
  \label{eq:entropy}
  C_{\text{GHZ}}(\psi) \leq \min_{\emptyset \neq I \subsetneq [m]} S(A_I)_\psi.
\end{equation}
Incidentally, the right hand side equals the minimum of $C_{\text{EPR}(i:j)}(\psi)$
over all $i\neq j$, according to Eq.~(\ref{eq:EPR-pairwise}), which even gives
an operational meaning to the bound, since from a GHZ-state between $m$ parties 
an EPR-state between any pair of parties can be obtained by LOCC. 

In the introduction we have already referenced several GHZ distillation protocols.
Here we briefly review a protocol based on \emph{entanglement combing} \cite{PhysRevLett.103.220501},
which results in a simple protocol and basic lower bound on the rate of 
GHZ distillation.
The following lemma is also going to be invoked in the proofs of our main results. 

\medskip
\begin{lemma}
\label{combing}
Let $\ket{\psi}^{B_{1}\ldots B_{m}}$ be a pure state shared among $m$ parties. 
The following rate of GHZ state is distillable from $\ket{\psi}$ under LOCC:
\begin{align}
  R_{\text{comb}} = \max_{i\in[m]}\left\{\min_{I\subseteq [m]\setminus i}\frac{S(B_{I})}{|I|}\right\}.
\end{align} 
In particular, if $\ket{\psi}$ is genuinely multi-party entangled (i.e. it is not 
a product state w.r.t. any bipartite cut), then $R_{\text{comb}} > 0$.
\end{lemma}

\begin{IEEEproof}
The entanglement combing protocol \cite{PhysRevLett.103.220501} turns the given 
state into bipartite entanglement shared between a distinguished party, say $i$, 
and each of the other parties $j \in [m]\setminus i$. 
Let $R_{j}$ denote the rate of the EPR pairs distilled between the 
distinguished party $B_{i}$ and another party $B_{j}$. 
The following rate region is proven optimal for this task:
\begin{align}
  \label{rcom}
  \forall I\subseteq [m]\setminus i \quad \sum_{j\in I }R_{j}\leq S(B_{I}).
\end{align}
By means of LOCC one can turn the combed entanglement into GHZ states shared between 
all parties. This can be done by letting party $i$ teleport their information using the 
EPR pairs. 
In this case, the rates have to be equal, i.e. $R_{1}=\ldots=R_{m} =: R_{\text{comb}}$. 
Thus, from the rate region for combing Eq.~(\ref{rcom}), we have as a necessary and
sufficent condition
\begin{align}
  \forall I\subseteq [m]\setminus i \quad |I|R_{\text{comb}}\leq S(B_{I}),
\end{align}
which is satisfied by $R_{\text{comb}} := \min_{I\subseteq [m]\setminus i}\frac{S(B_{I})}{|I|}$.
Finally, we optimise over the choice of distinguished party. 
\end{IEEEproof}

\medskip
\begin{remark}
The preceding result shows that unless the state is a product state across some 
bipartite cut, the GHZ-rate is always positive. 
Such states are called ``bi-separable'', in which case evidently no GHZ states 
can be distilled, cf. Eq.~(\ref{eq:entropy}).
The rate $R_{\text{comb}}$ is the baseline against which to compare any new protocol. 

It can be far from optimal, for example even if the initial 
$\ket{\psi}=\ket{\Gamma_m}$ is a GHZ state, then $R_{\text{comb}}=\frac{1}{m-1}$,
while obviously $C_{\text{GHZ}}(\Gamma_m) = 1$.
\end{remark}

\medskip
In the proofs of our GHZ distillation protocols, 
we shall use the following rules from the resource calculus of quantum
Shannon theory \cite{4626055}, where `$\geq$' means that the resources 
on the left hand side can be transformed asymptotically to the resources on the
right hand side by local operations only; $o$ is an arbitrarily small positive 
number. 

\medskip
\begin{lemma}[{Cancellation lemma \cite[Lemma 4.6]{4626055}}]
\label{cancellation}
For resources $\alpha, \beta, \gamma$, 
if $\alpha+\gamma\geq \beta+\gamma$, then $\alpha+o\gamma\geq \beta$.
\qed
\end{lemma}

\medskip
\begin{lemma}[{Removal of $o$ terms \cite[Lemma 4.5]{4626055}}]
\label{removal}
For resources, $\alpha, \beta, \gamma$, 
if $\alpha+o\gamma\geq \beta$ and $\alpha\geq z\gamma$ for some real $z>0$, 
then $\alpha\geq\beta$.
\qed
\end{lemma}

\medskip
\begin{theorem}[{Vrana and Christandl \cite[Thm.~1]{8678669}}]
\label{Vrana-Christandl}
Let $\ket{\psi}=\sum\psi_{x_{1}\ldots x_{m}}\ket{x_{[m]}}$ be a pure state written 
in the computational basis, and define $p(x_{1},\ldots,x_{m})=|\psi_{x_{1}\ldots x_{m}}|^{2}$,
the probability distribution of measuring $\psi$ in the computational bases locally. 
Define the region $\mathcal{R}_c$ as the set of rate tuples $R_{[m]}=(R_{1},\ldots,R_{m})$ 
satisfying the following conditions,
\begin{align}
  \label{coc}
  \forall I \subsetneq [m] \quad \sum_{j\in I}R_{j}\geq H(X_{I}|X_{[m]\setminus I}).
\end{align} 
Finally, let
$R_{\text{CO}}^{c} := \min_{R_{[m]}\in\mathcal{R}_{c}}\sum_{j=1}^{m}R_{j}$.
Then,
\[
  C_{\text{GHZ}}(\psi) \geq H(X_{[m]}) - R_{\text{CO}}^{c}.
\]
\end{theorem}

\begin{IEEEproof}
The $m$ terminals share $n$ copies of the pure state 
$|\psi\rangle=\sum_{x_{1}\ldots x_{m}}\psi_{x_{1}\cdots x_{m}}\ket{x_{1}}\cdots\ket{x_{m}}$,
i.e.
\begin{align*}
  \ket{\psi}^{\otimes n}           &= \sum_{x_{1}^{n}\ldots x_{m}^{n}}
                                       \psi_{x_{1}^{n}\ldots x_{m}^{n}}\ket{x_{1}^{n}}\cdots \ket{x_{m}^{n}}, \text{ where} \\
  \psi_{x_{1}^{n}\ldots x_{m}^{n}} &= \prod_{t=1}^{n}\psi_{x_{1,t}\cdots x_{m,t}} \text{ and}\\
  \ket{x^{n}_{j}}                  &= \ket{x_{j,1}}\otimes\cdots\otimes\ket{x_{j,n}}.
\end{align*}

Let $f_{j}:\mathcal{X}_j^n \rightarrow \mathcal{U}_j$ be the Slepian-Wolf 
hash function used by party $j$ in the classical 
part of the protocol of Theorem \ref{lowercr1} (omniscience), 
and $(\Delta^{(j,u_{[m]})}_{x_{[m]}^n}:x_{[m]}^n)$ the POVM (decision rule) 
that they use to recover $x_{[m]}^n$ when the classical messages 
$u_{[m]}$ are broadcast. 

In the first step, each party $j$ will apply an isometry based on the mappings 
$x_{j}^{n} \longmapsto (f_{j}(x_{j}^{n}),x_{j}^{n})$ for $j\in[m]$, namely
\begin{align*}
  V_j = \sum_{x_{j}^{n}} \ket{f_{j}(x_{j}^{n}),x_{j}^{n}}\!\bra{x_{j}^{n}},
\end{align*}
where $\ket{u}=\ket{f_{j}(x_{j}^{n})}$ are computational basis for some Hilbert 
space $U_j = \text{span}\{\ket{u} : u\in \mathcal{U}_j\}$.
The state at the end of the first step is 
\begin{align*}
  \ket{\psi'} = \sum_{x_{1}^{n}\ldots x_{m}^{n}} \psi_{x_{1}^{n}\ldots x_{m}^{n}}
                 \ket{x_{1}^{n},f_{1}(x_{1}^{n})}\cdots\ket{x_{m}^{n},f_{m}(x_{m}^{n})}.
\end{align*}
Next comes the coherent transmission of the hash value $u_j$ to other parties,
which in fact is implementing a multi-receiver cobit channel \cite{PhysRevLett.92.097902}, 
i.e. party $j$ wishes to implement the isometry
$\ket{u_j} \longmapsto \ket{u_j}^{\otimes m}$. 
This multi-receiver cobit channel can be implemented by teleportation
through GHZ states. In order to coherently transmit $nR_{j}$ bits, 
where $R_{j}\coloneqq \frac{1}{n}\log\abs{U_j}$, 
$nR_{j}$ GHZ states are needed, i.e. the following state:
\begin{align*}
  \ket{\Gamma_{m}}^{\otimes nR_{j}}
    =\left(\frac{1}{\sqrt{2}}(\ket{0}^{\otimes m}+\ket{1}^{\otimes m})\right)^{\otimes nR_{i}}.
\end{align*}
After implementing the multi-receiver cobit channel, the $j$-th party owns its initial 
share $\ket{x_{j}^{n}}$ as well as all the hash values broadcast to it. Thus, 
the overall state is
\begin{align*}
  \ket{\widetilde{\psi}} 
    &=\sum_{x_{1}^{n}\ldots x_{m}^{n}}\psi_{x_{1}^{n}\ldots x_{m}^{n}} 
        \ket{x_{1}^{n},f_{1}(x_{1}^{n})\ldots f_{m}(x_{m}^{n})} 
        \cdots \ket{x_{m}^{n},f_{1}(x_{1}^{n})\ldots f_{m}(x_{m}^{n})} .
\end{align*} 
Having received the hash values, each party proceeds to recovering $x_{[m]}^n$.
Each party locally runs its Slepian-Wolf decoder in a coherent fashion to work 
out the $\ket{x^{n}_{j}}$ of the other $m-1$ parties. 
More precisely, the $j$-th party applies the following controlled isometry 
on its corresponding systems:
\begin{align*}
 \sum_{u_{[m]}} \ketbra{u_{[m]}} \otimes V_{D}^{(j,u_{[m]})},
\end{align*}
where the coherent measurement isometry of the $j$-th party is defined as:
\begin{align}
 V_{D}^{(j,u_{[m]})}
   &=\sum_{\forall i\in [m]\,  x_i^n\in f_i^{-1}(u_i)} \sqrt{\Delta^{(j,u_{[m]})}_{x_{[m]}^n}} \otimes \ket{x_{[m]}^n},
\end{align}
with $\Delta^{(j,u_{[m]})}_{x_{[m]}^n}$ the POVM elements of the $j$-th decoder
acting on $A_j^n$. The classical result of Csisz\'ar and Narayan \cite{1362897}, 
i.e. Theorem \ref{lowercr1} in the diagonal case,
ensures successful decoding if the rates $R_{[m]}$ satisfy the conditions (\ref{coc}). 
The state after each party applied their decoding isometry is as follows:
\begin{align*}
  \ket{\overline{\psi}} = \sum_{x_{1}^{n}\ldots x_{m}^{n}} \psi_{x_{m}^{n}\ldots x_{m}^{n}}
    &\left( \sum_{\forall i\in [m]\, \xi_i^n\in f_i^{-1}(u_i)} 
               \sqrt{\Delta^{(1,u_{[m]})}_{\xi_{[m]}^n}} \ket{x_{1}^{n}}
                      \ket{f_{1}(x_{1}^{n})\ldots f_{m}(x_{m}^{n})} \ket{\xi_{[m]}^{n}} \right) \\
    &\otimes \cdots \\
    &\otimes\left( \sum_{\forall i\in [m]\, \xi_i^n\in f_i^{-1}(u_i)}
               \sqrt{\Delta^{(m,u_{[m]})}_{\xi_{[m]}^n}} \ket{x_{m}^{n}}
                      \ket{f_{1}(x_{1}^{n})\ldots f_{m}(x_{m}^{n})} \ket{\xi_{[m]}^{n}} \right).
\end{align*}

After decoding, by the coherent gentle measurement lemma \cite{796385,4544968}, 
the state will be $\sqrt{2m\varepsilon}$-close in trace distance to the following state:
\begin{align*}
\ket{\widehat{\psi}}
  &= \sum_{x_{[m]}^n} \psi_{x_{1}^{n}\ldots x^{n}_{m}}
                      \ket{x_1^n,f_{1}(x_{1}^{n})\ldots f_{m}(x^{n}_{m})} \ket{x_{[m]}^n}\\
  &\phantom{=========}
                      \otimes\cdots\otimes 
                      \ket{x_m^n,f_{1}(x_{1}^{n})\ldots f_{m}(x^{n}_{m})} \ket{x_{[m]}^n}.
\end{align*}
The details of the application of the coherent gentle measurement 
lemma are as follows. The coherent gentle measurement lemma ensures that for all parties $j\in [m]$
\begin{align*}
\sum_{\forall i\in [m]\,  x_i^n\in f_i^{-1}(u_i)}\sqrt{\Delta^{(j,u_{[m]})}_{x_{[m]}^n}}\ket{x_{j}^{n}}\otimes\ket{x_{[m]}^{n}}
\end{align*}
is $2\sqrt{\varepsilon(2-\varepsilon)}$ close in trace distance to $\ket{x_{j}^{n}}\otimes\ket{x_{[m]}^{n}}$ provided that the decoding error is not bigger than $\varepsilon$ Theorem \ref{lowercr1}. This implies
\begin{align*}
  \braket{\widehat{\psi}}{\overline{\psi}}
      &=\sum_{x_{1}^{n}\ldots x_{m}^{n}}|\psi_{x_{1}^{n}\ldots x_{m}^{n}}|^{2}
                \bra{x_{1}^{n}}\sqrt{\Delta^{(1,u_{[m]})}_{x_{[m]}^n}}\ket{x_{1}^{n}}
                \cdots \bra{x_{m}^{n}}\sqrt{\Delta^{(m,u_{[m]})}_{x_{[m]}^n}}\ket{x_{m}^{n}}\\
      &\geq \sum_{x_{1}^{n}\ldots x_{m}^{n}}|\psi_{x_{1}^{n}\ldots x_{m}^{n}}|^{2}
                \bra{x_{1}^{n}}\Delta^{(1,u_{[m]})}_{x_{[m]}^n}\ket{x_{1}^{n}} 
                \cdots \bra{x_{m}^{n}}\Delta^{(m,u_{[m]})}_{x_{[m]}^n}\ket{x_{m}^{n}}\\
      &\geq (1-\varepsilon)^{m} \geq 1-m\varepsilon.
\end{align*}
where the equality follows by substitution, the first inequality follows since 
$\sqrt{\Delta^{(m,u_{[m]})}_{x_{[m]}^n}}\geq\Delta^{(m,u_{[m]})}_{x_{[m]}^n}$
for $\Delta^{(m,u_{[m]})}_{x_{[m]}^n}\leq\mathbbm{1}$ and the second inequality follows from the assumption. 
Then, for the trace distance of pure states,
\begin{align*}
  \left\|\widehat{\psi}-\overline{\psi}\right\|_1 
     &=    2\sqrt{1-\left|\braket{\widehat{\psi}}{\overline{\psi}}\right|^{2}}\\
     &\leq 2\sqrt{1-(1-\varepsilon)^{2m}}
      \leq \sqrt{2m\varepsilon}.
\end{align*}

All parties now clean up their $U_{[m]}$-registers and their original $A_j^n$-register
by virtue of local unitaries, to arrive at the following state, up to trace norm
error $\sqrt{2m\varepsilon}$:
\begin{align}
\label{b-type-m}
 \ket{\widehat{\gamma}}
  &= \sum_{x_{[m]}^n} \psi_{x_{1}^{n}\ldots x^{n}_{m}}
                      \ket{x_{[m]}^n} \cdots \ket{x_{[m]}^n} 
\end{align}
To do that, note that the partial Slepian-Wolf isometries 
$V_j: \ket{x^{n}_{j}}\ket{0}^{E}\mapsto \ket{x^{n}_{j}}\ket{f_{j}(x^{n}_{j})}$ 
can be made a unitary by declaring 
$\ket{x^{n}_{j}}\ket{i}^{E}\mapsto \ket{x^{n}_{j}}\ket{i+f_{j}(x^{n}_{j})}$, 
where the addition is that of an abelian group on the ancillary 
register (e.g. integers modulo $|U_j|$). Once we have a unitary, 
the inverse is also a unitary, and can be applied locally. 

The above state can now be turned into a standard GHZ state at rate
$n H(X_{[m]})$ via the well-known entanglement concentration protocol,
just like the bipartite case \cite{PhysRevA.53.2046}. This involves 
measuring the \emph{type} $t$ of $x_{[m]}^n$, and noting that the 
phase and amplitude factors are constant along each type class, resulting in 
GHZ-type states after the measurement. To see that, let $\cT_{t}^{n}$ denote 
the set of sequences of the same type $t$, and let $\Pi_{t}$ be the projector 
onto the subspace spanned by $\cT_{t}^{n}$, i.e.
\begin{align*}
  \Pi_{t}=\sum_{x^{n}\in \cT_{t}^{n}}\ketbra{x^{n}}{x^{n}}.
\end{align*}
If the type resulting from the measurement does not belong to a typical type, 
then the protocol ends; with the properties of the type projectors, this happens with 
asymptotically small probability. Finally, we thus obtain approximately 
the following state resulting from the type class measurement (which 
is close to the initial state) 

\begin{align*}
  \frac{\Pi_{t}\otimes\cdots\otimes\Pi_{t}\ket{\widehat{\gamma}}}{\sqrt{p^n(\cT_{t}^{n})}}
    &= \sum_{x_{[m]}^n\in \cT_{t}^{n}} \sqrt{\widetilde{p}(x_{[m]}^{n})} \ket{x_{[m]}^n} \cdots \ket{x_{[m]}^n} \\
    &= \frac{1}{\sqrt{|\cT_{t}^{n}|}} \sum_{x_{[m]}^n\in \cT_{t}^{n}}\ket{x_{[m]}^n} \cdots \ket{x_{[m]}^n}, 
\end{align*}
where $p^n(\cT_{t}^{n})=\left|\bra{\widehat{\gamma}}\Pi_{t}\otimes\cdots\otimes\Pi_{t}\ket{\widehat{\gamma}}\right|$, 
$\widetilde{p}(x_{[m]}^{n})=\frac{p(x^{n}_{[m]})}{p^n(\cT_{t}^{n})}$ and 
$|\cT_{t}^{n}| \sim 2^{nH(X_{[m]})}$ for large $n$.

The protocol so far proves the following resource inequality:
\begin{align}
  \psi+R_{\text{CO}}[GHZ]+\infty[c\rightarrow c] \geq H(X_{[m]})[GHZ],
\end{align}
where $R_{\text{CO}}$ is the minimum of the sum of all rates of GHZ states used 
by parties to communication hash values. By using the Cancellation 
Lemma \ref{cancellation}, this implies now
\begin{align}
  \psi+o[GHZ]+\infty[c\rightarrow c] \geq (H(X_{[m]})\!-\!R_{\text{CO}})[GHZ].
\end{align}
In order to remove the $o$ term from the left-hand side of the resource inequality, 
we need Lemma \ref{removal}, which demands the following resource inequality to be true,
for some $\alpha > 0$:
\begin{align}
  \psi +\infty[c\rightarrow c] \geq \alpha [GHZ].
\end{align}  
Note that we need the asymptotic resource inequality, not some 
single-copy transformation (which might or might not imply the former), as 
prerequisite of the cancellation lemma. 
In Lemma \ref{combing} we have actually proven this inequality by virtue 
of entanglement combing. Therefore, we can remove the $o$ term and we have the result as we wished.
\end{IEEEproof}

\medskip
\begin{theorem}
\label{thm:GHZ-best}
Let $\ket{\psi}^{A_{1}\ldots A_{m}}$ be a pure state shared by $m$ spatially separated 
parties, and let $\mathcal{E}^{i}: A_{i}\rightarrow A_{i}\otimes X_{i}$ denote an 
instrument of party $i$, consisting of pure CP maps
$\mathcal{E}^i_x(\sigma) = E^i_x \sigma (E^i_x)^\dagger$ 
(which is why we may assume $A_i'=A_i$). 
Then, with the notation of Theorem \ref{lowercr2}, 
\begin{align*}
  C_{\text{GHZ}}(\psi) \geq H(X_{[m]}) - R_{\text{CO}}^{cq},
\end{align*}
where 
$R_{\text{CO}}^{cq}=\min_{\substack{R_{[m]}}\in\mathcal{R}_{cq}}\sum_{i=1}^{m} R_{i}$,
and $\mathcal{R}_{cq}$ is the rate region given as follows: 
\begin{align*}
  \forall j\in [m]\ \forall L\subseteq [m]\setminus j \quad 
      \sum_{i\in L} R_{i} \geq S(X_{L}|X_{[m]\setminus L}A_{j}').
\end{align*}
\end{theorem}

\begin{IEEEproof}
The proof follows from the techniques used in Theorem \ref{Vrana-Christandl}, and 
the result of Theorem \ref{lowercr2}:
making the protocol coherent and recycling.

Starting with a pure state, as in the proof of Theorem \ref{Vrana-Christandl}, each 
party applies their instrument coherently on its system, resulting in 
isometries $V_i:A_{i}\hookrightarrow A_{i}\otimes X_{i}$ defined as 
$V_i = \sum_{x\in\mathcal{X}_{i}} E^i_x \otimes \ket{x}$. 
The isometries act as follows on a single copy:
\begin{align*}
\ket{\widehat{\psi}}&= (V_{1}\otimes\cdots\otimes V_{m})\ket{\psi}^{A_{[m]}}\\
                    &= \sum_{x_{[m]}} (E^{1}_{x_{1}}\otimes\cdots\otimes E^{m}_{x_{m}})\ket{\psi}^{A_{[m]}}\otimes\ket{x_{[m]}}\\
                    &= \sum_{x_{[m]}} \sqrt{p(x_{[m]})}\ket{\widetilde{\psi}_{x_{[m]}}}^{A_{[m]}}\otimes\ket{x_{[m]}},
\end{align*}
where 
\begin{align*}
p(x_{[m]}) 
  = \bra{\psi}(E^{1}_{x_{1}}\otimes\cdots\otimes E^{m}_{x_{m}})^{\dagger}
              (E^{1}_{x_{1}}\otimes\cdots\otimes E^{m}_{x_{m}})\ket{\psi},
\end{align*}
and
\begin{align*}
\ket{\widetilde{\psi}_{x_{[m]}}}^{A_{[m]}}
  = \frac{(E^{1}_{x_{1}}\otimes\cdots\otimes E^{m}_{x_{m}})\ket{\psi}^{A_{[m]}}}{\sqrt{p(x_{[m]})}}.
\end{align*}

With $n$ copies of the initial pure state, we want to distill 
GHZ states from $n$ copies of $\ket{\widehat{\psi}}$, i.e.
\begin{align*}
  \ket{\widehat{\psi}}^{\otimes n}
    = \sum_{x_{[m]}^n} \sqrt{p^n(x_{[m]}^n)} \ket{x_{1}^n}\cdots\ket{x_{m}^n}
                                             \otimes\ket{\widetilde{\psi}_{x_{[m]}^n}}^{A_{[m]}^{n}},
\end{align*}
where $\ket{\widetilde{\psi}_{x_{[m]}^n}}^{A_{[m]}^{n}}$ is the quantum side information
at the disposal of the parties to help them with their decodings.

Similar to Theorem \ref{Vrana-Christandl}, 
in the first step each party coherently computes its hash value and broadcasts it 
coherently to the other parties via GHZ states. 
By applying the decoder of Theorem \ref{lowercr2} in a coherent fashion,
each party decodes $\ket{x_{[m]}^{n}}$ where the minimum rate of initial GHZ states is $R_{\text{CO}}^{cq}$. 
After the uncomputing of the hash value information and the local $X_j^n$, the state is approximately 
\begin{align}
\label{beforeEC}
  \ket{\widehat{\theta}}
    &= \sum_{x_{[m]}^n} \sqrt{p(x^{n}_{[m]})}\ket{x_{[m]}^n} \cdots \ket{x_{[m]}^n} \otimes \ket{\psi_{x_{[m]}^n}}^{A_{[m]}^n},
\end{align}
with residual states $\ket{\psi_{x_{[m]}^n}}$ on ${A_{[m]}^n}$. 
At the end, the parties implement the 
entanglement concentration protocol to get a standard GHZ state. That is, each one 
measures the joint type $t$ of $x_{[m]}^n$, i.e. they apply the projectors $\Pi_t$ 
from the proof of Theorem \ref{Vrana-Christandl}. If the result is a non-typical 
type, they abort the protocol; if it is typical, they proceed as follows to decouple 
the $A_{[m]}^n$-registers: all sequences $x_{[m]}^n$ from the type class $\cT_t^n$ are 
obtained by a permutation $\pi(x_{[m]}^n)\in S_n$ of a fiducial string 
$x_t^n \in \cT_t^n \subset \mathcal{X}_{[m]}^n$. The unitary $U_{\pi(x_{[m]}^n)}$
permuting the $n$ systems of $A_{[m]}^n$ do the same with a fiducial vector
$\ket{\psi_t} = \ket{\psi_{x_t^n}}$, i.e. $\ket{\psi_{x_{[m]}^n}} = U_{\pi(x_{[m]}^n)}\ket{\psi_t}$.
Party $j$ now applies the controlled permutation
\[
  U_j = \sum_{x_{[m]}^n \in \cT_t^n} \proj{x_{[m]}^n} \otimes (U_{\pi(x_{[m]}^n)})^{\dagger A_j^n}, 
\]
which maps the state to an approximation of
\begin{align}
\label{after-U-and-EC}
  \ket{\widetilde{\theta}}
    &= \frac{1}{\sqrt{|T_t^n|}} 
       \sum_{x_{[m]}^n \in \cT_t^n} \ket{x_{[m]}^n} \cdots \ket{x_{[m]}^n} \otimes \ket{\psi_t}^{A_{[m]}^n},
\end{align}
The last part, $\ket{\psi_t}^{A_{[m]}^n}$, is decoupled, as it only depends 
on $t$, and the remaining state is the desired GHZ state.
\end{IEEEproof}

\medskip
\begin{remark}
\label{rem:recycle}
The above protocol typically leaves some entanglement behind, in the 
form of the states $\ket{\psi_t}$. This entanglement could potentially be still useful for 
$m$-party GHZ distillation, but a more common situation is that it contains 
only entanglement between fewer ($\leq m-1$) parties, perhaps even only EPR states between 
a pair of parties. 

To distill it, essentially the same kind of protocol as in Theorem \ref{thm:GHZ-best}
can be applied, because $\ket{\psi_t} = \ket{\psi_{x_t^n}}$ is a product state 
across the $n$ $m$-partite systems, and by grouping identical states we can treat it
as a collection of i.i.d. states.
\end{remark}

\section{Conclusion}
\label{sec:end}
We have derived two achievability bounds for the distillable common randomness from 
mixed multipartite state and by making them coherent, we found two achievability 
bounds for the rate of GHZ distillation from a multipartite pure state. 
The first bound reproduces a recent result by Vrana and Christandl with genuinely
quantum Shannon theoretic methods, and the second improves on it in a truly quantum way. 

To our knowledge, it is the best currently known general bound. Note that it includes 
the lower bound from \cite{PhysRevA.72.052317},
which was formulated for tripartite state $\psi^{ABC}$, and is obtained by choosing 
a measurement basis $\{\ket{x}\}$ for $A$ and trivial 
(identity) instruments for $B$ and $C$ in Theorem \ref{thm:GHZ-best}; this gives 
a pure state decomposition $\psi^{BC}=\sum_x \lambda_x \ketbra{\psi_x}^{BC}$.
Let $\overline{E}_{BC}=\sum_x \lambda_x E(\ketbra{\psi_x})$ be the average bipartite
entanglement of the pure state decomposition. Define finally 
\begin{align*}
  \chi = \min{\{S(B),S(C)\}}-\overline{E}_{BC},
\end{align*} 
Then $\chi$ is an achievable rate of three-party GHZ distillation, but in addition
also EPR pairs between $B$ and $C$ at rate $\overline{E}_{BC}$ are distilled \cite{PhysRevA.72.052317}. 
This is consistent with our Theorem \ref{thm:GHZ-best} and 
Remark \ref{rem:recycle}, too: following through the proof, the leftover state,
there denoted $\ket{\psi_t}$, is precisely a tensor product of $\ket{\psi_x}$, 
with $x$ appearing $\sim n\lambda_x$ times.

\medskip
\begin{example}
Consider the three-qubit W-state
\begin{align*}
  \ket{W} = \frac{1}{\sqrt{3}} (\ket{001}+\ket{010}+\ket{100}). 
\end{align*}
Entanglement combing (Lemma \ref{combing}) results in a GHZ rate of
$R_{\text{comb}}=\frac12 H(\frac23,\frac13) \approx 0.4591$, but already the 
very simple yet ingenious protocol of \cite{FortescueLo} achieves $R_{FL}=0.5$, because
it extracts an EPR pair deterministically from every copy of the W-state, albeit
randomly distributed over the three possible pairs. 
Theorem \ref{Vrana-Christandl}, applied with the local computational bases, 
gets up to $R_{VC} = \log 3 - 1 \approx 0.585$. Namely, note that 
the omniscience information $X_1X_2X_3$ is jointly uniformly distributed 
over the set $\{001,010,100\}$, and so the conditions for communication for
omniscience in Theorem \ref{Vrana-Christandl} are $R_1 \geq H(X_1|X_2X_3) = 0$ and
cyclic, and $R_1+R_2 \geq H(X_1X_2|X_3) = \frac23$ and cyclic. 
Thus, $R_{\text{CO}}^c=\min R_1+R_2+R_3 = 3\cdot\frac12\cdot\frac23 = 1$.

The result from \cite{PhysRevA.72.052317} (recall that it is a 
special case of Theorem \ref{thm:GHZ-best}) however yields the 
seemingly very bad $R_{SVW} = \log 3 - \frac43 \approx 0.2516$, until 
we remember that as a bonus we get a rate of $\frac23$ of EPR states -- by the
symmetry of the W-state between any prescribed pair of parties, $AB$ or $BC$ or $AC$.
Pairs of these, from different pairs, can be fused to 
get an additional rate of $\frac13$ for GHZ generation, thus matching the total
of $R_{VC} = \log 3 - 1$.

We do not know, however, if this rate is optimal under general LOCC procedures, 
or even restricted to non-interactive communication protocols. 
\end{example}

\medskip
\begin{example}
Consider the tripartite fully antisymmetric state, also known as ``determinant state'', 
\begin{align*}
\ket{\alpha_3}=\frac{1}{\sqrt{6}} (\ket{123}+\ket{231}+\ket{312}-\ket{132}-\ket{213}-\ket{321}). 
\end{align*}
Similar to the previous example, we can evaluate the rate resulting from 
entanglement combing (Lemma \ref{combing}), 
$R_{\text{comb}}=\frac12 \log 3 \approx 0.7925$, because all three 
marginal qutrit states are maximally mixed. 
But Theorem \ref{Vrana-Christandl}, applied with the local computational bases, 
yields the much better $R_{VC}=\log 3 - \frac12 \approx 1.085$.
This is straightforward after realising that the computational bases measurements
result in the uniform distribution of $X_1X_2X_3$ over all $6$ permutations
$\{123,231,312,132,213,321\}$. The conditions for communication for
omniscience in Theorem \ref{Vrana-Christandl} are $R_1 \geq H(X_1|X_2X_3) = 0$ and
cyclic, and $R_1+R_2 \geq H(X_1X_2|X_3) = 1$ and cyclic. 
Thus, $R_{\text{CO}}^c=\min R_1+R_2+R_3 = \frac32$.
 
The result from \cite{PhysRevA.72.052317} gives the seemingly 
disappointing value $R_{SVW} = \log 3 - 1 \approx 0.585$; but as before, we
can salvage a rate of $1$ of EPR states between any prescribed pair of parties, 
thus contributing an additional rate of $\frac12$ for GHZ generation, once again 
matching the total of $R_{VC} = \log 3 - \frac12$.

Again, we do not know whether this is optimal, in particular whether there
is a better way of applying Theorem \ref{thm:GHZ-best}.
\end{example}

\medskip
\begin{example}
The \emph{flower state} \cite{HHHO:locking},
\[
  \ket{\varphi} = \frac{1}{\sqrt{2d}} \sum_{i=1}^d\sum_{j=0}^1 \ket{ij}^A \ket{ij}^B (H^j\ket{i})^C,
\]
where $H^{0}=\mathbbm{1}$ and $H^{1}$ is the $d$-dimensional quantum Fourier transform,
provides an example where Theorem \ref{thm:GHZ-best} is better 
than Vrana-Christandl. The former, by simply letting $A$ or $B$ measure 
and broadcast $j$, so that $C$ can undo the unitary $H^j$, yields the
clearly optimal $R_{SVW} = \log d$ (it is the local entropy of $C$, which 
is an upper bound on the distillable GHZ rate under arbitrary LOCC protocols). 

On the other hand, Theorem \ref{Vrana-Christandl} with the
computational bases for $A$ and $B$ (which seems like the evident choice, but we
have no full proof that it is optimal), and \emph{any} measurement of $C$, 
results in a rate $R_{VC} \leq \frac12\log d$.
This follows from Maassen-Uffink's entropic uncertainty relation \cite{MaassenUffink},
which reads as $I(X_1X_2;X_3) = I(X_1;X_3) \leq \frac12\log d$ (cf. \cite{HHHO:locking})
and some elementary algebraic manipulations. In detail, we have
$H(X_1)=H(X_2)=1+\log d$, and $H(X_3) \geq \log d$. On the other hand, the 
conditions for communication for omniscience in Theorem \ref{Vrana-Christandl} are 
$R_1 \geq H(X_1|X_2X_3) = 0$, $R_2 \geq H(X_2|X_1X_3) = 0$ and 
$R_3 \geq H(X_3|X_1X_2) \geq \frac12\log d$; furthermore
$R_1+R_2 \geq H(X_1X_2|X_3) \geq 1+\frac12\log d$, and the now redundant
$R_1+R_3 \geq H(X_1X_3|X_2) = H(X_3|X_2) \geq \frac12\log d$ and
$R_2+R_3 \geq H(X_2X_3|X_1) = H(X_3|X_1) \geq \frac12\log d$. 
Now for the net rate, we can reason
\[\begin{split}
  H(X_1X_2X_3) - (R_1+R_2+R_3) &\leq H(X_1X_2X_3) - H(X_3|X_1X_2) - H(X_1X_2|X_3) \\
                               &=    H(X_1X_2) - H(X_1X_2|X_3) \\
                               &=    I(X_1X_2;X_3) \leq \frac12\log d
\end{split}\]
using the lower bounds for $R_3$ and $R_1+R_2$ in the first line, the chain rule
for the entropy in the second line, and finally the entropic uncertainty relation. 
\end{example}

\bigskip
In future work we are going to apply the machinery developed in this paper to 
secret key distillation against an adversary who is initially correlated and 
eavesdrops on the public classical communication between the parties, and to the
distillation of GHZ states from \emph{mixed} initial states. Regarding the former, 
we can quite evidently apply Theorem \ref{lowercr2} to a general state
$\rho^{A_1\ldots A_m E}$ and local instruments $\mathcal{E}_i:A_i\rightarrow A_i'X_i$, 
to first attain omniscience $X_{[m]}$ at all legal parties, and then hashing 
this information down using privacy amplification \cite{privacy-amplification}, 
resulting in a lower bound
\[
  C_S(\rho) \geq S(X_{[m]}|E) - R_{\text{CO}}^{cq}
\]
on the distillable secret key. 
Regarding GHZ distillation, we would apply these protocols to a purification 
$\ket{\psi}^{A_1\ldots A_m E}$ of $\rho^{A_1\ldots A_m}$, and for pure 
instruments as in Theorem \ref{thm:GHZ-best} we expect to obtain the 
lower bound
\[
  C_{\text{GHZ}}(\rho) \geq S(X_{[m]}|E) - R_{\text{CO}}^{cq}
\]
on the distillable GHZ rate. This will require a generalization of the techniques 
from \cite{2005RSPSA.461..207D} to the multi-party setting with non-interactive
communication, of turning a privacy amplification step into a decoupling procedure. 

Furthermore, note that we have focused our attention on non-interactive 
protocols, but it seems evident that in general there is an 
advantage in protocols using interactive communication, 
i.e. of fully general CR distillation, cf. \cite{interactive,5508611}.
In this context it is an important question to determine which class of 
interactive communication protocols, when applied to a quantum state, 
can be made coherent and thus yields achievable rates for GHZ distillation.

\medskip\noindent
\textbf{Acknowledgments.} 
AW thanks Clarice Starling for illuminating discussions regarding 
several aspects of public and secret information.
The authors acknowledge financial support 
by the Baidu-UAB collaborative project ``Learning of Quantum 
Hidden Markov Models'', the Spanish MINECO (project FIS2016-86681-P) 
with the support of FEDER funds, and the Generalitat de Catalunya
(project 2017-SGR-1127). FS also supported by the Catalan Government
001-P-001644 QuantumCAT within the ERDF Program of Catalunya.

\appendix

\section*{Classical correlated source coding with side information at the decoder}
\label{app:simultaneous-SW}

The analysis of multi-party common randomness distillation via our 
omniscience protocol leads quite naturally to the consideration of 
classical source coding with quantum side information at the 
decoder \cite{PhysRevA.68.042301,WINTER:PHD}. Here we present the 
necessary definitions, and prove a new coding theorem for achieving all 
points of the rate region directly by random binning and a quantum joint 
typicality decoder, rather than successive decoding and time sharing as 
in the cited previous works. 

\medskip
A multipartite correlated classical-quantum (cq-)source is given by a cq-state 
\begin{equation}
  \label{eq:multiple-source}
  \rho^{X_{1}\ldots X_{k}B} = \sum_{x_{[k]}} p(x_{[k]}) \proj{x_1}^{X_1}\ox\cdots\ox\proj{x_k}^{X_k} \ox \rho_{x_{[k]}}^B,
\end{equation}
where $X_i$ (which we can identify with a classical random variable)
is observed by the $i$-th encoder, who sends a function of $X_i$ to the
decoder. The decoder has the quantum system $B$ and by measuring it depending 
on all the messages received from the $k$ encoders attempts to reconstruct
$X_{[k]}$ with high probability. 

\medskip
\begin{definition}
\label{multiple-source}
An $n$-block coding scheme with quantum side information at the decoder  
for the cq-source $\rho^{X_{[k]}B}$ consists of $k$ encoding functions 
$f_{i}:\cX^{n}_{i} \rightarrow [M_i]$ and decoding POVMs $\Lambda^{(\mu_{[k]})}$ on 
$B^n$, one for each $\mu_{[k]} = \mu_1\ldots\mu_k \in [M_1]\times\cdots\times[M_k]$, 
and indexed by $\cX_{1}^{n}\otimes\cdots\otimes \cX_{k}^{n}$. 
Its rates are the numbers $\frac1n \log M_i$, and
its average error probability is
\begin{align*}
  P_{e} := 1 - \sum_{x_{[k]}^n} p^n(x_{[k]}^n) 
                                \Tr \rho_{x_{[k]}^n}^{B^n}\Lambda^{\bigl(f_{[k]}(x_{[k]}^n)\bigr)}_{x_{[k]}^n}.
\end{align*}
Here, $f_{[k]}(x_{[k]}^n) = f_1(x_1^n)\ldots f_k(x_k^n)$ is the $k$-tuple 
of compressed data. 

A $k$-tuple $(R_1,\ldots,R_k)$ is called an achievable rate tuple if there 
exists $n$-block coding schemes for all $n$, such that its error probability 
converges to zero, $P_e \rightarrow 0$, and the rates $\frac1n \log |M_i|$ 
converge to $R_i$. 
The set of achievable rate tuples is called the rate region of the compression 
problem described by $\rho^{X_{[k]}B}$.
\end{definition}

\medskip
By definition, the rate region is a closed subset of the positive orthant $\RR_{\geq 0}^k$, 
that is closed under increasing individual vector components. By the time sharing 
principle, it is also convex. 
Necessary and sufficient conditions for the rate region were proved 
in \cite[{Thm.~IV.14 \& Cor.~IV.16}]{WINTER:PHD}, which are the ones 
expected from Slepian-Wolf coding: 
\begin{equation}
  \label{eq:cq-coding-region}
  \forall\ I\subseteq [k] \quad \sum_{i\in I} R_i \geq S(X_I|X_{[k]\setminus I}B).
\end{equation}
While the necessity of these conditions is rather straightforward, we will 
be concerned here with their sufficiency.  
In the cited PhD thesis, this is obtained by showing that the extreme points 
of the polytope (\ref{eq:cq-coding-region}) can be achieved, which in turn  
is done by successive decoding of the $j$-th sender's information $X_j^n$, in 
an order given by the extreme point in question, of which there are $k!$, one for 
each permutation of the parties $[k]$.
The rest follows by the convexity and openness-above of the rate region. 

The following lemma shows that it is possible to construct a code 
by random binning and with a simultaneous decoding scheme 
that achieves directly every point in the rate region. This is essential in applications, 
such as ours, where there are multiple decoders with different side informations 
for the same compressed data. 

\medskip
\begin{lemma}[{Simultaneous quantum decoder}]
\label{lemma:cq-compression-simultaneous}
With the above notation, suppose the rates $R_{i}=\frac{1}{n}\log|M_{i}|$ satisfy 
the following inequalities for some $\delta>0$,
\begin{align*}
  \forall\ \emptyset\neq I\subseteq [k] \quad \sum_{i\in I} R_i \geq S(X_I|X_{[k]\setminus I}B) + \delta,
\end{align*}
where the entropies are with respect to the state (\ref{eq:multiple-source}). 

Then, for independent $2$-universal random functions $F_i:\cX_i^n \rightarrow [M_i]$,
there exists simultaneous decoding POVMs $\bigl(\Lambda^{(\mu_{[k]})}_{x^{n}_{[k]}}\bigr)$ 
such that the expectation of the average error probability over all codes converges to
zero: $\EE_{F_1\ldots F_k} P_e \rightarrow 0$, as $n\rightarrow\infty$.
\end{lemma}

\begin{IEEEproof}
We will use Sen's construction of jointly typical POVM elements \cite[Sec. 5]{Sen:jointly-typical}, 
which is stated as Lemma \ref{lemma:Sen} below, in the simplified form in which we need it.

Consider $\left(\rho^{X_{[k]}B}\right)^{\ox n} = \rho^{X_{[k]}^nB^n}$ and for the 
RVs $X_1^n, \ldots, X_k^n$ denote the set of jointly entropy-typical sequences
by $\cT$. This means that $\Pr\{X_{[k]}^n\in\cT\} \geq 1-\eta \rightarrow 1$ 
as $n\rightarrow\infty$ and that for every $x_{[k]}^n\in\cT$ and all $I\subseteq[k]$,
\[
  2^{-nH(X_I)-n\beta} \leq p^n(x_I^n) \leq 2^{-nH(X_I)+n\beta}, 
\]
with an arbitrarily chosen $\beta>0$. 

Next we apply Lemma \ref{lemma:Sen} to the $(k+1)$-party state $\rho^{X_{[k]}^nB^n}$ to obtain first 
an ``augmented'' state $\rho^{X_{[k]}^nB^n} \ox \tau^{C^n}$ for a suitable system $C$ and a universal 
state $\tau^C$ (actually the maximally mixed state), where we think of $BC$ as a new 
quantum system $\widetilde{B}$, so that the augmented state is still a $(k+1)$-party 
cq-state. Note that $\tau^C$ can be created locally at $B$.
Lemma \ref{lemma:Sen} then gives us an approximation $\widetilde{\rho}^{X_1^n \ldots X_k^n \widetilde{B}^n}$
and a POVM element $E$ with the properties stated in the lemma. 
Importantly, both this state and the POVM element share the original cq-structure:
\begin{align*}
  \widetilde{\rho}^{X_1^n \ldots X_k^n \widetilde{B}^n}
    &=  \sum_{x_{[k]}^n} p^n(x_{[k]}^n) \proj{x_{[k]}^n}^{X_{[k]}^n} \ox \widetilde{\rho}_{x_{[k]}^n}^{\widetilde{B}^n}, \\
  E &= \sum_{x_{[k]}^n} \proj{x_{[k]}^n}^{X_{[k]}^n} \ox E_{x_{[k]}^n}.
\end{align*}
By restricting the latter to typical $x_{[k]}^n$, we obtain 
\[
  E' := \sum_{x_{[k]}^n\in\cT} \proj{x_{[k]}^n}^{X_{[k]}^n} \ox E_{x_{[k]}^n},
\]
which does not affect property 1 in Lemma \ref{lemma:Sen}, and preserves
property 3, while property 2 becomes the only slightly worse
$\Tr \widetilde{\rho}^{X_{[k]}^n\widetilde{B}^n} E' \geq 1-2\gamma-\gamma'-\eta$.

Finally, for the encoding by independent $2$-universal functions $F_j$, after 
the receiver obtains $\mu_1\ldots \mu_k$, we need a decoding POVM for recovering 
$x_{[k]}^n \in \cT \cap F_1^{-1}(\mu_1)\times\cdots\times F_k^{-1}(\mu_k)$
from $\rho_{x_{[k]}^n}^{B^n} \ox \tau^{C^n}$.
We use the square-root measurement $(\Lambda_{x_{[k]}^n})$ constructed from the $E_{x_{[k]}^n}$,
$x_{[k]}^n \in \cT \cap F_{[k]}^{-1}(\mu_{[k]})$: 
\[
  \Lambda_{x_{[k]}^n} 
     = \left(\sum_{x_{[k]}^{\prime n} \in \cT\cap{F}_{[k]}^{-1}({F}_{[k]}(x_{[k]}^n))} E_{x_{[k]}^{\prime n}}\right)^{-\frac12}
                             E_{x_{[k]}^n}
       \left(\sum_{x_{[k]}^{\prime n} \in \cT\cap{F}_{[k]}^{-1}({F}_{[k]}(x_{[k]}^n))} E_{x_{[k]}^{\prime n}}\right)^{-\frac12}.
\]
To upper bound its error probability, we employ the Hayashi-Nagaoka operator 
inequality, stated below as Lemma \ref{lemma:HN}:
\[\begin{split}
  P_e &\leq 1-p^n(\cT) 
             + \sum_{x_{[k]}^n\in\cT} p^n(x_{[k]}^n) \Tr(\rho_{x_{[k]}^n}^{B^n}\ox\tau^{C^n})\Lambda_{x_{[k]}^n} \\
      &\leq \eta + \gamma 
             + \sum_{x_{[k]}^n\in\cT} p^n(x_{[k]}^n) \Tr\widetilde{\rho}_{x_{[k]}^n}^{\widetilde{B}^n}\Lambda_{x_{[k]}^n} \\
      &\leq \eta + \gamma
             + \sum_{x_{[k]}^n\in\cT} p^n(x_{[k]}^n) 
                        \left( 2\,\Tr\widetilde{\rho}_{x_{[k]}^n}^{\widetilde{B}^n}(\1-E_{x_{[k]}^n})
                               +  4\sum_{x_{[k]}^{\prime n} \in \cT\cap{F}_{[k]}^{-1}({F}_{[k]}(x_{[k]}^n))
                                                                                               \setminus x_{[k]}^n} 
                                          \Tr\widetilde{\rho}_{x_{[k]}^n}^{\widetilde{B}^n}E_{x_{[k]}^{\prime n}} \right) \\
      &\leq \eta + 5\gamma + 2\gamma'
             + 4 \sum_{x_{[k]}^{\prime n}\in\cT} 
                        \Tr E_{x_{[k]}^{\prime n}}
                            \left( \sum_{x_{[k]}^n \in \cT\cap{F}_{[k]}^{-1}({F}_{[k]}(x_{[k]}^{\prime n}))
                                                                                            \setminus x_{[k]}^{\prime n}}
                                                                p^n(x_{[k]}^n) \widetilde{\rho}_{x_{[k]}^n} \right),
\end{split}\]
where in the first line we declare an error for non-typical $x_{[k]}^n$, and in the second line 
have used property 1 in Lemma \ref{lemma:Sen}; in the third line, we used Lemma \ref{lemma:HN},
applied to $T = E_{x_{[k]}^n}$ and 
$S = \sum_{x_{[k]}^{\prime n} \in \cT\cap F_{[k]}^{-1}(F_{[k]}(x_{[k]}^n))\setminus x_{[k]}^n} E_{x_{[k]}^{\prime n}}$;
finally, in the fourth line we use property 2 in Lemma \ref{lemma:Sen} for the first term 
in the bracket, and for the second term simply reorganise the double sum.

Thus, to bound the expected error probability, over the random choice of the $F_j$, 
we need a bound on the expected state in the round brackets in the last line 
of the above chain of inequalities. To do so, we distinguish the different cases 
of coordinates $\emptyset\neq I\subseteq[k]$ in which $x_{[k]}^n$ and $x_{[k]}^{\prime n}$ differ:
\[\begin{split}
  \EE_{{F}_{[k]}} \left( \sum_{x_{[k]}^n \in \cT\cap{F}_{[k]}^{-1}({F}_{[k]}(x_{[k]}^{\prime n}))\setminus x_{[k]}^{\prime n}}
                                                                           p^n(x_{[k]}^n) \widetilde{\rho}_{x_{[k]}^n} \right) 
                      &\leq \sum_{\emptyset\neq I\subseteq[k]} 
                               \frac{1}{\prod_{i\in I} M_i} 
                                  \sum_{x_{[k]}^n\in\cT \atop \text{s.t. } x_{I^c}^n=x_{I^c}^{\prime n}}
                                                          p^n(x_{[k]}^n) \widetilde{\rho}_{x_{[k]}^n} \\
                      &=: \sum_{\emptyset\neq I\subseteq[k]} 
                               \frac{1}{\prod_{i\in I} M_i} p(x_{I^c}^{\prime n}) \widetilde{\rho}_{x_{I^c}^{\prime n}},
\end{split}\]
with the shorthand notation $I^c = [k]\setminus I$ for the set complement.
Furthermore, in the first line we have used the $2$-universality of the $F_j$, as well 
as their independence, and in the second line note that the probabilities and states
$p(x_{I^c}^{\prime n}) \widetilde{\rho}_{x_{I^c}^{\prime n}}$ appear in the marginal
\[
  \widetilde{\rho}^{X^n_{I^c}\widetilde{B}^n} 
      = \sum_{x^n_{I^c}} p(x^n_{I^c}) \proj{x^n_{I^c}}^{X^n_{I^c}} \ox \widetilde{\rho}_{x^n_{I^c}}^{\widetilde{B}^n}.
\]
This means that 
\[\begin{split}
  \EE_{{F}_{[k]}} P_e 
                 &\leq \eta + 5\gamma + 2\gamma' 
                        + 4 \sum_{\emptyset\neq I\subseteq[k]} \frac{1}{\prod_{i\in I} M_i}
                            \sum_{x_{[k]}^{\prime n}\in\cT} 
                                    \Tr\, p(x_{I^c}^{\prime n}) \widetilde{\rho}_{x_{I^c}^{\prime n}} E_{x_{[k]}^{\prime n}} \\
                 &\leq \eta + 5\gamma + 2\gamma' 
                        + 4 \sum_{\emptyset\neq I\subseteq[k]} \frac{2^{nH(X_I)+n\beta}}{\prod_{i\in I} M_i}
                            \sum_{x_{[k]}^{\prime n}\in\cT} 
                                    \Tr\, p(x_{I}^{\prime n})p(x_{I^c}^{\prime n}) 
                                          \widetilde{\rho}_{x_{I^c}^{\prime n}} E_{x_{[k]}^{\prime n}} \\
                 &=    \eta + 5\gamma + 2\gamma' 
                        + 4 \sum_{\emptyset\neq I\subseteq[k]} \frac{2^{nH(X_I)+n\beta}}{\prod_{i\in I} M_i}
                                    \Tr\left(\widetilde{\rho}^{X_I^n}\ox\widetilde{\rho}^{X_{I^c}^n\widetilde{B}^n}\right)E' \\
                 &\leq \eta + 5\gamma + 2\gamma' 
                        +  4\sum_{\emptyset\neq I\subseteq[k]} \frac{2^{nH(X_I)+n\beta}}{\prod_{i\in I} M_i}
                                    2^{-D_h^\epsilon\bigl(\rho^{X_{[k]}^nB^n}\big\|\rho^{X_I^n}\ox\rho^{X_{I^c}^nB^n}\bigr)} \\
                 &\leq \eta + 5\gamma + 2\gamma' 
                        +  4\sum_{\emptyset\neq I\subseteq[k]} \frac{2^{nH(X_I)+n\beta}}{\prod_{i\in I} M_i}
                                                                            2^{-nI(X_I:X_{I^c}B)+n\beta}                \\
                 &\leq \eta + 5\gamma + 2\gamma' 
                        + 4 \sum_{\emptyset\neq I\subseteq[k]} 2^{n\left(H(X_I|X_{I^c}B)+2\beta-\sum_{i\in I} R_i\right)},
\end{split}\]
where in the second line we use entropy typicality of the $x_{[k]}^{\prime n}$; to get
the third line simply insert the forms of $\widetilde{\rho}$ and $E$ above; in the 
fourth line we use property 3 in Lemma \ref{lemma:Sen}, and in the fifth 
we invoke the asymptotic equipartition property (AEP) for the hypothesis 
testing relative entropy, stated as Lemma \ref{lemma:AEP} below.

Hence, choosing $\beta = \delta/3$, we obtain as an upper bound on the expected error 
probability 
$\EE_{{F}_{[k]}} P_e \leq \eta + 5\gamma + 2\gamma' + 2^{k+2} 2^{-n\delta/3}$, 
which converges to $0$ as $n\rightarrow\infty$ (and $\epsilon \rightarrow 0$ sufficiently slowly).
\end{IEEEproof}

\medskip
Here follow the technical lemmas from the literature invoked in the proof. 
\medskip
\begin{lemma}[{Sen's jointly typical operators \cite[{Lemma 1 in Sec. 5, cf. Sec. 1.3}]{Sen:jointly-typical}}]
\label{lemma:Sen}
Let $X_1\ox\cdots\ox X_k\ox B$ be a $(k+1)$-partite classical-quantum system with 
finite-dimensional classical system $X_i$ and a finite-dimensional quantum system $B$, 
and $\epsilon>0$. Then there exists a Hilbert space $C$ and a state $\tau^C$ 
on it such that for any cq-state $\sigma^{X_1\ldots X_kB}$, there is a
cq-state $\widetilde{\sigma}^{X_1\ldots X_k\widetilde{B}}$ and a POVM element 
$E$ (also of cq-form) on $X_1\ldots X_k\widetilde{B}$, where $\widetilde{B} = B \ox C$, 
with the following properties:
\begin{enumerate}
  \item $\frac12 \left\| \widetilde{\sigma}^{X_{[k]}\widetilde{B}} - \sigma^{X_{[k]}B}\ox\tau^C \right\|_1 \leq \gamma$, 
  \item $\Tr \widetilde{\sigma}^{X_{[k]}\widetilde{B}} E \geq 1-2\gamma-\gamma'$,
  \item for all $\emptyset \neq I \subseteq [k]$, 
        $\Tr \left(\widetilde{\sigma}^{X_I}\ox\widetilde{\sigma}^{X_{[k]\setminus I}\widetilde{B}}\right)E 
         \leq 2^{-D_h^\epsilon\left(\sigma^{X_{[k]}B}\big\|\sigma^{X_I}\ox\sigma^{X_{[k]\setminus I}B}\right)}$.
\end{enumerate}
Here, $\gamma = \sqrt{2}^{k+1} \sqrt[4]{\epsilon}$ 
and $\gamma' = 2^{k+2^{k+5}}\sqrt{\epsilon}$.
\hfill\IEEEQED
\end{lemma}

\medskip
\begin{lemma}[{Hayashi and Nagaoka \cite{HayashiNagaoka}}]
\label{lemma:HN}
For a POVM element $0\leq T\leq \1$ and a positive operator $S>0$,
\[
  \phantom{====================}
  \1-(S+T)^{-\frac12}T(S+T)^{-\frac12} \leq 2(\1-T) + 4S.
  \phantom{====================}\IEEEQED
\]
\end{lemma}

\medskip
\begin{lemma}[{Hiai and Petz \cite{HiaiPetz:Stein}; Ogawa and Nagaoka \cite{OgawaNagaoka:Stein}}]
\label{lemma:AEP}
For any two states $\rho$ and $\sigma$, and $0 < \epsilon < 1$,
\[
  \lim_{n\rightarrow\infty} \frac1n D_h^\epsilon\bigl(\rho^{\otimes n}\big\|\sigma^{\otimes n}\bigr) = D(\rho\|\sigma),
\]
where $D(\rho\|\sigma) = \Tr\rho(\log\rho-\log\sigma)$ is the Umegaki quantum
relative entropy. 
\hfill\IEEEQED
\end{lemma}

\bigskip
Using the joint decoder for independent random binning we obtain a new proof for
the achievability of the rate region (\ref{eq:cq-coding-region}) for correlated 
classical source coding with quantum side information at the 
decoder \cite[{Thm.~IV.14 \& Cor.~IV.16}]{WINTER:PHD}, which does away with the 
successive decoding of the different parts of the source. This detail allows the 
solution of a more demanding problem that was out of reach of the methods 
in \cite{WINTER:PHD}, correlated source coding for multiple decoders with 
quantum side information. Rather than giving the formal definition, let us 
just indicate the changes to Definition \ref{multiple-source}: the source is 
given by a cq-state 
\begin{equation}
  \label{eq:cqq-source}
  \rho^{X_{[k]}B_{[q]}} 
      = \sum_{x_{[k]}} p(x_{[k]}) \proj{x_1}^{X_1}\ox\cdots\ox\proj{x_k}^{X_k} \ox \rho_{x_{[k]}}^{B_{[q]}}
\end{equation}
with $q$ quantum systems $B_1,\ldots, B_q$. A block code for this 
system is still given by encoding function $f_i$ for each user $i\in[k]$,
such that $\mu_i = f_i(x_i^n)$ is broadcast to all $q$ decoders;
but now we need a decoding POVM $\Lambda^{(j;\mu_{[k]})}$ on $B_j^n$ 
for each decoder $j\in[q]$ that satisfy all the decoding error probability 
criterion for the cq-source $\rho^{X_{[k]}B_j}$. The random binning protocol
of Lemma \ref{lemma:cq-compression-simultaneous} then shows that the region
\begin{equation}
  \label{eq:cqq-coding-region}
  \forall\ j\in[q]\ \forall\ I\subseteq [k] \quad \sum_{i\in I} R_i \geq S(X_I|X_{[k]\setminus I}B_j)
\end{equation}
is achievable for rates at which all decoders can successfully decode $X_{[k]}$
simultaneously. That the above conditions are necessary is also evident, so 
Eq. (\ref{eq:cqq-coding-region}) is precisely the rate region. 

In \cite{Sen:jointly-typical,Sen:jointly-typical-apps} it was shown that the 
joint typicality Lemma \ref{lemma:Sen} leads to simultaneous, joint-typicality 
decoders for the classical-quantum multiple access channel (cq-MAC), in fact 
essentially optimal one-shot bounds. Using a well-known reduction of MAC to 
Slepian-Wolf, we can also derive the iid rate region from the present result
Eq. (\ref{eq:cqq-coding-region}), even in the presence of multiple receivers, 
cf.~\cite{Ahlswede:2-2}. 
Namely, for the $k$-sender, $q$-receiver cq-MAC that takes input
$x_{[k]}=x_1\ldots x_k$ to $\rho_{x_{[k]}}^{B_{[q]}}$, and in the simplest case a 
product distribution $p(x_{[k]}) = p_1(x_1)\cdots p_k(x_k)$, consider the cq-state 
as in Eq. (\ref{eq:cqq-source}). For block length $n$ and the random code as in 
Lemma \ref{lemma:cq-compression-simultaneous}, consider the bins restricted
to the typical sequences, for sender $i$ this is $\cT_i$, the sequences
typical for the probability distribution $p_i$, and denote their respective 
cardinalities by $N_i = 2^{nR_i'}$. Then, we have with high probability that
most of the bins are good codes for all decoders and that for most of the
bins in turn $\bigl|R_i' - (H(X_i)-R_i)\bigr| \leq \frac{1}{k}\delta$, 
and so $\sum_{i\in I} R_i' \leq {\displaystyle \min_j}\; I(X_I:B_j|X_{[k]\setminus I}) - 2\delta$
for all $\emptyset \neq I \subseteq [k]$. For any rate tuple satisfying these constraints 
there exists thus asymptotically good codes. 

To get the full rate region, we also need an auxiliary random variable $U$ such that 
$X_1,\ldots,X_k$ are independent conditionally on $U$; 
then, every tuple of rates $R_i'$ such that
\[
  \forall\ I\subseteq[k]\quad \sum_{i\in I} R_i' \leq \min_j I(X_I:B_j|X_{[k]\setminus I}U), 
\]
is asymptotically achievable for transmitting $k$ independent messages from the 
separate senders to all receivers $B_j$, $j\in[q]$. The proof is quite similar to 
the sketch above and is omitted.

\bibliographystyle{IEEEtran}


\end{document}